\documentclass[twocolumn,aps,prl,showpacs,superscriptaddress,floatfix]{revtex4}

\usepackage[dvips]{color} 
\usepackage{graphicx} 
\usepackage{bm}

\begin{document}

\title{Liquid Drop Stability of a Superdeformed Prolate \\Semi-Spheroidal 
Atomic Cluster}

\author{D. N. Poenaru}
\email[]{poenaru@fias.uni-frankfurt.de}

\affiliation{
Horia Hulubei National Institute of Physics and Nuclear
Engineering, \\P.O. Box MG-6, RO-077125 Bucharest-Magurele, Romania}
\affiliation{Frankfurt Institute for Advanced Studies,
J. W. Goethe Universit\"at, 
        Max-von-Laue-Str. 1,    D-60438 Frankfurt am Main,   Germany}

\author{R. A. Gherghescu}

\affiliation{
Horia Hulubei National Institute of Physics and Nuclear
Engineering, \\P.O. Box MG-6, RO-077125 Bucharest-Magurele, Romania}
\affiliation{Frankfurt Institute for Advanced Studies,
J. W. Goethe Universit\"at, 
        Max-von-Laue-Str. 1,    D-60438 Frankfurt am Main,   Germany}

\author{A. V. Solov'yov}
\affiliation{Frankfurt Institute for Advanced Studies,
J. W. Goethe Universit\"at, 
        Max-von-Laue-Str. 1,    D-60438 Frankfurt am Main,   Germany}

\author{W. Greiner}
\affiliation{Frankfurt Institute for Advanced Studies,
J. W. Goethe Universit\"at, 
        Max-von-Laue-Str. 1,    D-60438 Frankfurt am Main,   Germany}

\date{\today}

\begin{abstract}
Analytical relationships for the surface and curvature energies of oblate
and prolate semi-spheroidal atomic clusters have been obtained. By modifying
the cluster shape from a spheroid to a semi-spheroid (including the flat
surface of the end cup) the most stable shape was changed from a sphere to a
superdeformed prolate semi-spheroid. Potential energy surfaces vs.
deformation and the number of atoms, $N$, illustrate this property
independent of $N$.

\end{abstract}


\maketitle

The density functional theory \cite{par89b} is successfully employed in the
field of atomic cluster physics. Alternatively, with less computational
effort, one can use as a first approximation some simple models (see the
reviews  \cite{bra93rmp,hee00ib} and the references therein)
replacing the many-body effects by an effective single-particle potential,
since to a good approximation the delocalized conduction electrons of
neutral small metallic clusters form a Fermi liquid like the atomic nucleus
\cite{sch92b}.

The liquid drop model (LDM) \cite{ray78plms} dominated for many decades the
theory of nuclear fission, starting with the first explanation, given in
1939 by Lise Meitner and O. Frisch, of the induced fission process
discovered by O. Hahn and F. Strassmann. One of the most cited paper in the
field was published in the same year \cite{boh39pr}. We have used the LDM to
develop the analytical superasymmetric fission model allowing 
to predict in 1980 heavy particle radioactivity \cite{ps84sjpn80}. 

In 1990 W. A. Saunders \cite{sau90prl} adapted the LDM to the atomic cluster
physics, and explained the increase of fissionability with decreasing size
of the charged metal cluster which was observed in experiments.  The
following year
J. P. Perdew et al. \cite{per91prl} presented a LDM for a neutral metal
cluster with $z=1, 2, 3, 4$ valence electrons. They mentioned in the
abstract that the LDM ``originally developed for finite systems (nuclei),
may actually be more appropiate for infinite ones (metals).'' For the ground
state properties of neutral clusters or the fission of doubly or multiply
charged clusters, the LDM \cite{yan99mc,sau92pra,koi93zpd,bra89prb,nah97pr}
expresses the smooth part of the total energy to which the shell corrections
\cite{str67np,kni84prl,cle85prb,yan93prb,bre94prb,yan95prb,mar96pr,yan97prl}
may be added. The interplay between LDM, shape deformations and
Strutinsky shell corrections (including the case of fission) for free
clusters was reviewed by C. Yannouleas {\em et al.} \cite{yan99mc}.

An interesting application of the liquid drop surface tension for producing
micro or nanoscale objects was recently reported \cite{pyc07prl}.
We believe that LDM completed with shell corrections \cite{str67np} 
could be very useful for the physics of the atomic
clusters deposited on different types of surfaces. It may provide very
rapidly a first solution which can be eventually refined within density
functional theory. 

The nanostructured coating of surfaces by cluster deposition 
\cite{bar05n,chi06prl} is at present a rapidly growing field.
By analyzing some shapes of cluster deposited on a surface obtained by using
scanning probe microscopy \cite{see99apl,bon05no}, one can see that a
semi-spheroid with the $z$ axis of cylindrical symmetry oriented
perpendicularly on the surface plane may be a good approximation.
We present analytical results for the deformation-dependent surface and
curvature energies, which has the advantage of shortest computer time and
easiest interpretation.

In all LDM studies published until now, both in nuclear and atomic cluster
physics, the most stable shape (minimum deformation energy) was a sphere.
One would expect that for a semi-spheroid the most stable shape would be a
semi-sphere or an oblate shape. For the first time, we found the surprising
result that the superdeformed prolate semi-spheroid is in this case the
equilibrium shape.

We investigate the stability of semi-spheroidal shapes by assuming, as a
first approximation of one possibility which can be met in practice, a
vanishing interaction energy with the surface on which the cluster is
deposited, so that the neutral atomic cluster may be considered to be free.
Other types of shapes obtained from a spheroid by removing more or less than
its half \cite{sem07} will be considered in the future; the interaction with
surface will be taken into account as well. Our choice is motivated by the
fact that the corresponding shell model \cite{p275arxiv} allows to obtain
analytical formulae for the single-particle energies. The remarkable result
of this shell model is that for the first time the maximum degeneracy is
associated to the superdeformed prolate shape. In this way both LDM and the
shell structure suggest the enhanced stability of such a distorted shape.

We are using the standard notation of $(\rho, z)$ for the axially symmetric
dimensionless cylindrical coordinates. 
When the shape is a semi-spheroid the length
scale is given by the radius of a sphere with the same volume, 
$R_s=2^{1/3}R_0=2^{1/3}r_sN^{1/3}$,
in which $N$ is the number of atoms, $r_s$ is the Wigner-Seitz radius
(2.117~{\AA} for Na \cite{bra89prb,yan95prb,sol03prl}) and  $\rho=\rho
(z)$ is the surface equation given by 
\begin{equation} \rho^2=\left \{
\begin{array}{ll} (a/c)^2(c^2 - z^2) & \
\ \ \ z \geq 0 \\ 0 & \ \ \ \ z < 0 \end{array} \right .
\end{equation}
where $a$ is the minor (major) semiaxis for prolate (oblate)
semi-spheroid and $c$ is the major (minor) semiaxis for prolate (oblate)
semi-spheroid. Volume conservation leads to 
$a^2c=1$. 

It is convenient to choose the deformation \cite{cle85prb} parameter 
$\delta$ defined by
\begin{equation}
a=\left(\frac{2-\delta}{2+\delta}\right)^{1/3} \; \; ; \; \;
c=\left(\frac{2+\delta}{2-\delta}\right)^{2/3} 
\label{clem}
\end{equation}
so that
\begin{equation}
\frac{a}{c}=\frac{2-\delta}{2+\delta}=a^3 \; \; ; \; \;
c=\frac{1}{a^2}
\label{clem2}
\end{equation}
\begin{figure}[htb]
\centerline{\includegraphics[width=8cm]{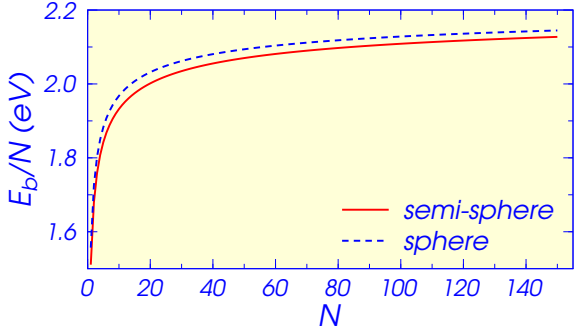}} 
\caption{
Binding energy per atom ($-E_b/N$) versus the number of atoms $N$
for Na clusters. Comparison between spherical and semi-spherical shapes.
\label{bese1}}
\end{figure}
The eccentricity is defined by the equation
\begin{equation}
e^2=\left \{ \begin{array}{ll}
1-a^2/c^2 & prolate \ \ (a<c) \\ 
a^2/c^2 -1 & oblate \ \ \ \ (a>c) \end{array}       \right .
\end{equation}

The deformation energy with respect to semi-spherical shape of a neutral
cluster is expressed as
\begin{eqnarray}
E-E^{s0} & =(E_s-E_s^{s0})+(E_c-E_c^{s0}) \nonumber \\    &
=E_s^{s0}\left( \frac{E_s}{E_s^{s0}} -
1 \right) + E_c^{s0}\left( \frac{E_c}{E_c^{s0}} -1 \right)
\end{eqnarray}
\begin{equation}
E-E^{s0}=E_s^{s0}\left( B_{surf}^s - 1 \right) + 
E_c^{s0}\left( B_{curv}^s -1 \right)
\end{equation}
where for a semi-sphere one has
\begin{equation}
E_s^{s0}=3\pi R_s^2\sigma = 3\pi 4^{1/3}R_0^2\sigma=\frac{3}{4^{2/3}}E_s^0
\end{equation}
\begin{equation}
E_c^{s0}=2\pi R_s\gamma_c=2\pi 2^{1/3}R_0 \gamma_c=\frac{E_{curv}^0}{4^{1/3}}
\end{equation}
with $E_s^0$ and $E_{curv}^0$ the surface and curvature energy of a sphere 
with the same volume.
When the atom is electrically charged, one has to add the Coulomb energy.

Clusters and nuclei are ``leptodermous'' systems characterized by a constant
density in the volume and a thin surface layer allowing to expand their
binding energy in terms of powers of $N^{1/3}$ (see a detailed discussion
in refs. \cite{bra93rmp,fra01arnps}). Despite the fact that this expansion
is {\it a priori} valid only for large enough systems ({\it e.g.}
Na$_{2654}$) it \cite{per91prl} 
``predicts the energy per electron $E/N$ accurately (within
0.03~eV) even for $N=1$''. 
For a spherical Na cluster \cite{bra89prb,yan95prb} the 
binding energy, in eV, is given by
\begin{equation}
E_N=-2.252N+0.541N^{2/3}+0.154N^{1/3}
\label{bind}
\end{equation}
where $E_v^0=-2.252N$~eV is proportional to the volume (assumed to be
conserved), 
$E_s^0=0.541N^{2/3}$~eV is proportional to the surface area
and the surface tension $\sigma$
\begin{equation}
E_s^0=4\pi R_0^2\sigma = 4\pi r_s^2\sigma N^{2/3}
\end{equation}
hence $4\pi r_s^2\sigma = 0.541$~eV for Na clusters. The curvature energy
$E_{curv}^0=0.154N^{1/3}$~eV is proportional to the integrated curvature
ant to the curvature tension $\gamma_c$
\begin{equation}
E_{curv}^0=4\pi R_0 \gamma_c = 4\pi r_s \gamma_c N^{1/3}
\end{equation}
where $4\pi r_s\gamma_c = 0.154$~eV for Na clusters.

The numerical coefficients in eq. (\ref{bind}) have been determined
\cite{bra89prb,yan95prb} by fitting the extended Thomas-Fermi local density
approximation total energy \cite{eka84prb} for spherical shapes. In the
fig.~2 of ref.~\cite{bra89prb} the smooth line expressed by eq.~(\ref{bind})
is compared to the dots from ref.~\cite{eka84prb}. The two sets of data
coincide at magic numbers; shell effects, explaining the deformation energy
of non-spherical atomic clusters, may be added by using Strutinsky's
\cite{str67np} procedure.

The liquid drop part (volume, surface, and curvature terms)
of the binding energy of Na semi-spherical clusters will
be in eV:
\begin{eqnarray}
E_{sN}=-2.252N & +\frac{3}{4^{2/3}}0.541N^{2/3} \nonumber \\    &
+\frac{1}{4^{1/3}}0.154N^{1/3}
\label{binds}
\end{eqnarray}
A cluster with a spherical shape is more tightly bound than a cluster with
a semi-spherical shape, as shown in figure~\ref{bese1}.

The area of a surface of revolution about $z$ axis is
\begin{equation}
S=\int dS = 2\pi R_0^2
\int_{-c}^{+c}\rho(z)\sqrt{1+\left(\frac{d\rho}{dz}\right)^2}dz
\end{equation}
The deformation-dependent surface energy for cylindrical symmetry
$B_{surf}=S/S_0=S/4\pi R_0^2$ is given by
\begin{equation}
B_{surf}=\frac{1}{2}\int_{-c}^{+c}dz\rho\sqrt{1+\rho^{\prime
2}}=\frac{1}{2}\int_{-c}^{+c}dz\sqrt{\rho^2+(\rho\rho^\prime)^2}
\end{equation}
which for spheroidal (oblate or prolate) shapes becomes
\begin{eqnarray}
B_{surf} &
=\frac{a}{2}\int_{-c}^{+c}dz\sqrt{1+z^2\left(\frac{a^2}{c^2}-
1\right)\frac{1}{c^2}} \nonumber \\ &
=a/(2c^2)\int_{-c}^{+c}dz\sqrt{c^4+z^2(a^2-c^2)}
\label{bs}
\end{eqnarray}
because for a spheroid
\begin{equation}
\rho^2=\frac{a^2}{c^2}(c^2-z^2) \ \ ; \ \ 
\rho\rho^\prime=-\frac{a^2}{c^2}z
\end{equation}
and
\begin{equation}
\rho^\prime\rho^\prime + \rho\rho^{\prime \prime}=-\frac{a^2}{c^2} \ \ ; \ \ \ 
\rho\rho^{\prime
\prime}=-\frac{a^2}{c^2}\left(1+\frac{a^2z^2}{c^2\rho^2}\right)
\end{equation}

The local curvature 
\begin{equation}
\kappa=0.5({\cal{R}}^{-1}_1+{\cal{R}}^{-1}_2)
\end{equation}
of a sphere is $1/R_0$ and the integrated curvature 
\begin{equation}
K=\int dS \kappa
\end{equation}
is $4\pi R_0$.
${\cal{R}}_1$ and ${\cal{R}}_2$ are the two principal radii of
curvature at a local point on the surface.
\begin{figure}[htb]
\centerline{\includegraphics[width=8cm]{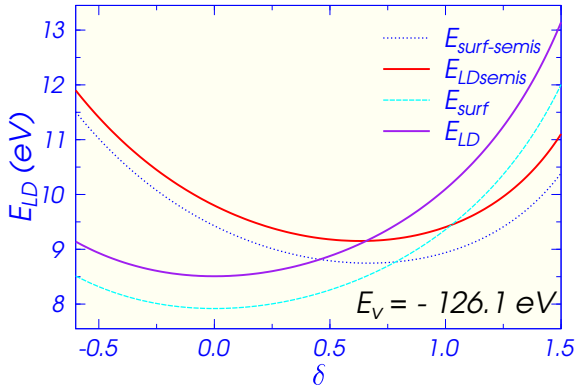}} 
\caption{Absolute values of the deformation-dependent
liquid drop surface and total (surface plus curvature) energy
versus the deformation parameter $\delta$
for Na$_{56}$ clusters. Comparison between spheroidal and semi-spheroidal shapes.
The volume energy $E_v=-126.1$~eV was not included in $E_{LD}$ and
$E_{LDsemis}$.
\label{lds+}}
\end{figure}
The shape-dependent part of the curvature energy for cylindrical symmetry is
given by
\begin{equation}
B_{curv}=\frac{1}{4\pi R_0}\int dS\kappa=
\frac{2\pi R_0^2}{4\pi R_0}\int_{-c}^{+c}dz\kappa\rho\sqrt{1+\rho^{\prime 2}}
\end{equation}
The principal radii of curvature of a shape with axial symmetry
\cite{p260np05} are expressed as
\begin{equation}
{\cal{R}}_1=R_0\rho\sqrt{1+\rho^{\prime 2}} \ \ \ \ \ \
{\cal{R}}_2=-R_0\frac{(1+\rho^{\prime 2})^{3/2}}{\rho^{\prime \prime}}
\end{equation}
hence
\begin{equation}
B_{curv}=\frac{1}{4}\int_{-c}^{+c}dz\frac{1+\rho^{\prime 2}-\rho\rho^{\prime
\prime}}{1+\rho^{\prime 2}}
\end{equation}

For the relative surface energy of a semi-sphere we obtain
\begin{eqnarray}
B_{surf}(\delta=0) & 
=(4\pi R_s^2/2+\pi R_s^2)/(4\pi  R_0^2) \nonumber \\ & =(3/4)
R_s^2/R_0^2=3/4^{2/3} 
\end{eqnarray}
and the corresponding curvature energy
\begin{eqnarray}
B_{curv}(0) & =K/4\pi R_0=(4\pi R_s/2)/(4\pi R_0) \nonumber \\ &
=(1/2)(R_s / R_0)=1/ 4^{1/3} 
\end{eqnarray}

We give $\rho, z, a, c$ in units of $R_s=2^{1/3}R_0$, hence
according to the volume conservation,
$a^2cR_s^3/2=R_0^3$ so that $a^2c=1$
\begin{equation}
B_{surf}^s=\frac{S}{3\pi R_s^2}= \frac{a^2}{3} +
\frac{2}{3}\int_0^c dz\sqrt{\rho^2+(\rho
\rho^\prime)^2}
\end{equation}
\begin{equation}
B_{surf}^s=\frac{a^2}{3}+\frac{2a}{3c^2}\int_0^c
dz\sqrt{c^4+z^2(a^2-c^2})
\end{equation}
because for a semi-spheroid there is a plane circular surface 
with the area $\pi R_s^2a^2$. The deformation dependent curvature energy of
a semi-spheroid, $B_{curv}^s=K/(2\pi R_s)$ is
\begin{equation}
B_{curv}^s=\frac{1}{2\pi R_s}\int dS\kappa =\frac{R_s^2}{R_s}
\int_0^c dz \kappa \sqrt{\rho ^2+ (\rho \rho ^\prime)^2}
\end{equation}

The integrated curvature for the plane surface is zero, hence
\begin{eqnarray}
B_{curv}^s & =\frac{1}{2}\int_0^cdz\left(1-\frac{\rho \rho ^{\prime
\prime}}{1+\rho^{\prime 2}} \right) \nonumber \\ & 
= \frac{c}{2} + \frac{a^2c^2}{2}\int_0^c
\frac{dz}{c^4+z^2(a^2-c^2)}
\end{eqnarray}
and the second term of 
this equation can be simplified by taking into account that $a^2c=1$.

When $a>c$ (oblate semi-spheroid) and $e^2=a^2/c^2-1$, $a^2-c^2=e^2c^2$, and
\begin{eqnarray}
B_{surf}^s & =\frac{a^2}{3} + \frac{2a}{3c^2}\int_0^cdz\sqrt{c^4+z^2c^2e^2}
\nonumber \\ & =
\frac{a^2}{3} + \frac{2a}{3c^2}ec\int_0^cdz\sqrt{\frac{c^2}{e^2} +
z^2}=
\end{eqnarray}
\begin{equation}
B_{surf}^s=\frac{a}{3}\left[2a+\frac{c}{e}\ln
\left(e+\frac{a}{c}\right)\right]
\end{equation}
\begin{eqnarray}
B_{curv}^s & =\frac{c}{2} + \frac{a^2c^2}{2}\int_0^c\frac{dz}{c^4+z^2c^2e^2}
\nonumber \\ & =
\frac{c}{2} + \frac{a^2c^2}{2c^2e^2}\int_0^c\frac{dz}{z^2+\frac{c^2}{e^2}}
\end{eqnarray}
\begin{equation}
B_{curv}^s=\frac{c}{2} + \frac{1}{2ce^2}\left(\frac{e}{c}\arctan e \right)
=\frac{c}{2} +\frac{a^2}{2ce}\arctan e
\end{equation}
For a semi-spherical shape $B_{surf}^s=B_{curv}^s=1$.

Asymptotically for $a\gg c$ one has $e \rightarrow a/c$ and consequently
\begin{equation}
\lim_{a \gg c}  B_{surf}^s \rightarrow \frac{2a^2}{3} \ \ ; \ \ \
\lim_{a \gg c}  B_{curv}^s \rightarrow \frac{\pi a}{4}
\end{equation}
\begin{figure}[htb]
\includegraphics[width=5.5cm]{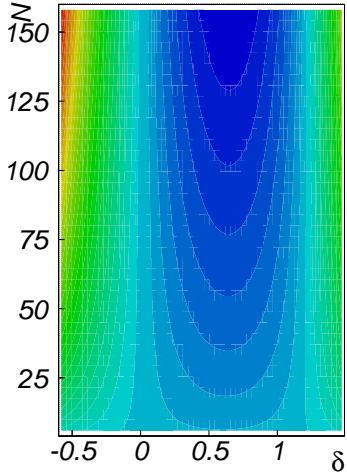} 
\caption{
Liquid drop deformation energy contour plot for Na clusters vs.
deformation $\delta$ and the number of atoms, $N$. 
Deformation energy relative to the
semi-spherical shape.
\label{ldcont}}
\end{figure}

When $c>a$ (prolate semi-spheroid), 
from the definition of the eccentricity we have $c^2-a^2=e^2c^2$,
hence
\begin{eqnarray}
B_{surf}^s & =\frac{a^2}{3} + \frac{2a}{3c^2}\int_0^c dz \sqrt{c^4-e^2c^2z^2}
\nonumber \\ & 
=\frac{2a^2}{3} + \frac{ac}{3e}\arcsin e = \frac{a}{3}\left( 2a+\frac{c}{e}
\arcsin e \right)
\end{eqnarray}
\begin{eqnarray}
B_{curv}^s & =\frac{c}{2} + \frac{a^2c^2}{2}\int_0^c \frac{dz}{c^4-e^2c^2z^2}
\nonumber \\ & = \frac{c}{2}+ \frac{a^2}{4ce}\ln \left | \frac{1+e}{1-e} \right |
\end{eqnarray}
Similar equations for spheroidal shapes may be found in Ref. \cite{ber61pr},
where the notation $\eta=a/c$ is used.

Asymptotically for $c\gg a$ one has $e \rightarrow 1$ and consequently
\begin{equation}
\lim_{c \gg a}  B_{surf}^s \rightarrow \frac{\pi ac}{6} \ \ ; \ \
\lim_{c \gg a}  B_{curv}^s \rightarrow \frac{c}{2}
\end{equation}

A comparison between the two LDM deformation energies in figure~\ref{lds+}
illustrates the essential difference between the spheroidal and
semi-spheroidal shapes of a Na cluster with 56 atoms; 
in the former case the most stable configuration is
a sphere, but in the latter it is a superdeformed prolate semi-spheroid with 
$\delta = 0.65$ ($c/a=1.96$). 

The contour plot of deformation energy vs. the deformation $\delta$ and the
number of atoms $N$ from figure \ref{ldcont} shows the general trend of
stabilty of superdeformed prolate shapes independent of the number of atoms
in the cluster.

Consequently by modifying the shape from a spheroid to a semi-spheroid the
most stable shape was changed from a sphere to a superdeformed prolate
semi-spheroid. In applications it will not be easy to measure the
interaction energy of the deposited cluster with the surface. 
The geometry may differ according to the strength of interaction between the
cluster and the substrate. By observing
in the experiment a prolate semi-spheroidal-like shape one can conclude that
the interaction energy is very small.

\begin{acknowledgments}
This work was partly supported by Deutsche Forschungsgemeinschaft, Bonn,
and by Ministry of Education and Research, Bucharest.
\end{acknowledgments}


\end{document}